\documentclass[superscriptaddress]{revtex4}
\usepackage[dvips]{graphicx}
\usepackage{epsfig}
\usepackage{amsmath}
\usepackage{amssymb}
\usepackage{amsthm}
\usepackage{calc}

\newtheorem{theorem}{Theorem}

\newtheorem{lemma}{Lemma}

\newtheorem{definition}{Definition}
\newtheorem{proposition}{Proposition}

\newcommand{\remove}[1]{}

\begin{document}

\title{Trends Prediction Using Social Diffusion Models}

\author{Yaniv Altshuler, Wei Pan, Alex (``Sandy'') Pentland \\ 
{\ } \\ MIT Media Lab \\
\texttt{\{yanival,panwei,sandy\}@media.mit.edu}}

\begin{abstract}
The importance of the ability of predict trends in social media has been growing rapidly in the past few years with the growing dominance of social media in our everyday's life. Whereas many works focus on the detection of anomalies in networks, there exist little theoretical work on the prediction of the likelihood of anomalous network pattern to globally spread and become ``trends''.
In this work we present an analytic model the social diffusion dynamics of spreading network patterns. Our proposed method is based on information diffusion models, and is capable of predicting future trends based on the analysis of past social interactions between the community's members.
We present an analytic lower bound for the probability that emerging trends would successful spread through the network. We demonstrate our model using two comprehensive social datasets --- the \emph{Friends and Family} experiment that was held in MIT for over a year, where the complete activity of 130 users was analyzed, and a financial dataset containing the \mbox{complete} activities of over 1.5 million members of the \emph{eToro} social trading community.
\end{abstract}

\maketitle

\section{Introduction}

We live in the age of social computing. Social networks are everywhere, exponentially increasing in volume, and changing everything about our lives, the way we do business, and how we understand ourselves and the world around us. The challenges and opportunities residing in the social oriented ecosystem have overtaken the scientific, financial, and popular discourse.

In this paper we study the evolution of trend spreading dynamics in social networks. Where there have been numerous works studying the topic of anomaly detection in networks (social, and others), literature still lacks a theoretic model capable of predicting \emph{how do network anomalies evolve}. When do they spread and develop into global trends, and when they are merely statistical \mbox{phenomena}, local fads that get quickly forgotten?
We give an analytically proven lower bound for the spreading probability, capable of detecting \mbox{``future trends'' -- spreading} behavior patterns that are likely to become prominent trends in the social \mbox{network}.

We demonstrate our model using social networks from two different domains. The first is the \emph{Friends and Family} experiment \cite{Aharony2011}, held in MIT for over a year, where the complete activity of 130 users was analyzed, including data concerning their calls, SMS, MMS, GPS location, accelerometer, web activity, social media activities, and more. The second dataset contains the complete financial transactions of the \emph{eToro community} members -- the world's largest ``social trading'' platform, allowing users to trade in currency, commodities and indices by selectively copying trading activities of prominent traders.

The rest of the paper is organized as follows~: Section \ref{sec.related_work} discusses related works. The information diffusion model is presented in Section \ref{sec.problem} and its applicability is demonstrated in Section \ref{sec.results}, and concluding remarks \mbox{are given in Section \ref{sec.conc}.}

%

\section{Related Work}
\label{sec.related_work}

\noindent \textbf{Diffusion Optimization}.
Analyzing the spreading of information has long been the central focus in the study of
social networks for the last decade~\cite{huberman2009social}~\cite{leskovec2009meme}. Researchers
have explored both the offline networks structure by asking and incentivizing
users to forward real mails and E-mails~\cite{dodds2003experimental},
and online networks by collecting and analyzing data from various sources such as \emph{Twitter} feeds~\cite{kwak2010twitter}.

\remove{
Researchers believe that this line of works can help understand the influence of
individuals in the modern WWW, which is fully equipped with all types of social media websites
and tools~\cite{cha2010measuring}, and it can eventually lead to accurate prediction and active optimization and
construction for successful and low-cost viral market campaigns such as the DARPA Challenge~\cite{pickard2011time}.
However,the information diffusion process on social networks
is overwhelmingly complicated: the outcome is clearly sensitive to many parameters and model
settings that are not entirely well understood and modeled correctly.
As a result, accurate prediction and optimization for promoting a certain trend still remains as an active topic.
}

The dramatic effect of the network topology on the dynamics of information diffusion in communities was demonstrated in works such as~\cite{choi2010role}~\cite{nicosia2011impact}. One of the main challenges associated with modeling of behavioral dynamics in social communities stems from the fact that it often involves stochastic generative processes.
While simulations on realizations from these models can help us
explore the properties of networks~\cite{herrero2004ising}, a theoretical analysis is much more
appealing and robust. In this work we present results are based on a pure theoretical analysis.

The identity and composition of an initial ``seed group'' in trends analysis has also been the topic of much research.
Kempe et al. applied theoretical analysis on the seeds selection problem~\cite{kempe2003maximizing} based on two simple adoption models:
\emph{Linear Threshold Model} and \emph{Independent Cascade Model}.
Recently, Zaman et al. developed a method to trace rumors back in the topological spreading path to identify sources
in a social network~\cite{shah2009rumors}, and suggest such method can be used to
locate influencers in a network. Some scholars express their doubts and concerns for
the influencer-driven viral marketing
approach, suggesting that ``everyone is an influencer''~\cite{bakshy2011everyone},
and companies ``should not rely on it''~\cite{watts2007viral}. They argue that the
content of the message is also important in determining its spreads, and likely the adoption
model we were using is not a good representation for the reality.

Our work, on the other hand, focuses on predicting emerging trends given a current snapshot of the network and
adoption status, rather than finding the most influential nodes. We provide a lower bound for the probability that an emerging trend would spread throughout the network, based on the the analysis of the diffusion process outreach, which is largely missing in current literature.

\noindent \textbf{Adoption Model}.
A fundamental building block in trends prediction that is not yet entirely clear to scholars
is the adoption model, modeling individuals' behavior based on the social signals they are exposed to. Centola has shown both theoretical
and empirically that a complex contagion model is indeed more precise for diffusion~\cite{centola2007complex,centola2010spread}.
Different adoption models can dramatically alter the model outcome~\cite{dodds2004universal}.
In fact, a recent work on studying mobile application diffusions using mobile
phones demonstrated that in real world the diffusion process
is a far more complicated phenomenon, and a more realistic model was proposed in ~\cite{funfaaai}.
Our results also incorporate this realistic diffusion model.

\remove{
\subsection{Social Influence}

The topic of social influence and how it can be inferred based on constrained information derived from online domains has been the topic of many works~\cite{onnela:2010facebookpnas}.
Much research concerning the prediction of users' behavior based on the dynamics in their community has been carried out in the past, using a variety of approaches such as sociological methods \cite{game-prediction-socio-HSU,friedman1-secondlife}, communities-oriented approaches \cite{game-prediction-communities-HSU}, game theory \cite{game-prediction-gametheory-CESABIANCHI} and various machine learning methods \cite{game-prediction-doppelganger-ORWANT}.
Still, to date there has been no attempts to generalize works regarding local influence models into a network-oriented prediction model.
}

\noindent \textbf{Trends Prediction and Our Proposed Model}.
In this work we study the following question~:
Given a snapshot of a social network with some behavior occurrences (i.e. an emerging trend), what is the probability that these occurrences (seeds) will result in a viral diffusion and a wide-spread trend (or alternatively, dissolve into oblivious). Though this is similar
to the initial seed selection problem~\cite{kempe2003maximizing}, we believe that the key factor to succeed in a viral marketing campaign optimization is a better analytical model for the diffusion process itself.

The main innovation of our model is the fact that it is based on a fully analytical framework
with a scale-free network model. Therefore, we manage to overcome the dependence on
simulations for diffusion processes that characterizes most of the works in this field~\cite{choi2010role,banerjee2004reaction}. We are able to do so by decomposing the diffusion process to the transitive random walk of ``exposure agents'' and the local adoption model based
on ~\cite{funfaaai}.
While there are some works that analyze scale-free network~\cite{meloni2009traffic} most of them come short to providing accurate results, due to the fact that they calculate the expected values of the global behaviors dynamics. Due to strong ``network effect'' however, many real world networks display much less coherent patterns, involving local fluctuations and high variance in observed parameters, rendering such methods highly inaccurate and sometimes impractical.
Our analysis on the other hand tackle this problem by modeling the diffusion process on scale-free networks in a way which takes into account such interferences, and can bound their overall effect on the network.

\section{Trend Prediction in Social Networks}
\label{sec.problem}
One of the main difficulty of trends-prediction stems from the fact that the first spreading phase of ``soon to be global trends'' demonstrate significant similarity to other types of anomalous network patterns. In other words, given several observed anomalies in a social network, it is very hard to predict which of them would result in a wide-spread trend and which will quickly dissolve into oblivious.

We model the community, or social network, as a graph $G$, that is comprised of $V$ (the community's members) and $E$ (social links among them). We use $n$ to denote the size of the network, namely $|V|$. In this network, we are interested in predicting the future behavior of some observed anomalous pattern $a$. Notice that $a$ can refer to a growing use of some new web service such as \emph{Groupon}, or alternatively a behavior such as associating oneself with the ``\emph{99\% movement}''.

Notice that ``exposures'' to trends are transitive. Namely, an ``exposing'' user generates ``exposure agents'' which can be transmitted on the network's social links to ``exposed users'', which can in turn transmit them onwards to their friends, and so on. We therefore model trend's exposure interactions as movements of random walking agents in a network. Every user that was exposed to a trend $a$ generates $\beta$ such agents, on average.

We assume that our network is (or can be approximated by) a scale free network $G(n,c,\gamma)$, namely, a network of $n$ users where the probability that user $v$ has $d$ neighbors follows a power law~:
\[
P(d) \sim c \cdot d^{-\gamma}
\]

We also define the following properties of the network~:
\begin{definition}
Let $V_{a}(t)$ denote the group of network members that at time $t$ advocate the behavior associated with the potential trend $a$.
\end{definition}

\begin{definition}
Let us denote by $\beta > 0$ the average ``\emph{diffusion factor}'' of a trend $a$. Namely, the average number of friends a user who have been exposed to the trend will be talking about the trend with (or exposing the trend in other ways).
\end{definition}

\begin{definition}
Let $P_{\Delta}$ be defined as the probability that two arbitrary members of the network vertices have degrees ratio of $\Delta$ or higher~:
\[
P_{\Delta} \triangleq Prob \left[ deg(u) > \Delta \cdot deg(v) \right]
\]
\end{definition}

\begin{definition}
\label{def.prob.wxpoaw.min}
\remove{For any $t$ and $\Delta_{t}$} We denote by \remove{$\sigma_{+}$ and } $\sigma_{-}$ the \remove{highest and } ``\emph{low temporal resistance}'' \mbox{of the network~:}
\remove{
\[
\sigma_{+} \triangleq \min \left\{ \begin{array}{c}
                                     1 \leq d \leq 2 \\
                                     1 \leq \Delta
                                   \end{array}
 \ \left| \
1 - e^{-\Delta \cdot d \cdot \frac{\beta^{\Delta_{t}} \cdot |V_{a}(t)}{n}} \cdot
\left(1 - c \cdot \frac{1-\frac{d-1}{d}^{\gamma-1}}{\gamma-1}\right) \cdot (1 - P_{\Delta})\right.\right\}
\]
}
\[
\forall t, \Delta_{t} \qquad , \qquad
\sigma_{-} \triangleq \max \left\{  1 \leq \Delta
 \ \left| \
1 - e^{-\Delta \cdot \frac{\beta^{\Delta_{t}} \cdot |V_{a}(t)|}{n}} \cdot
(1 - P_{\Delta})\right.\right\}
\]
\end{definition}

\begin{definition}
Let $P_{Local-Adopt}(a,v,t,\Delta_{t})$ denote the probability that at time $t + \Delta_{t}$ the user $v$ had adopted trend $a$ (for some values of $t$ and $\Delta_{t}$). This probability may be different for each user, and may depend on properties such as the network's topology, past interactions between members, etc.
\end{definition}

\begin{definition}
Let $P_{Local}$ denote that expected value of the local adoption probability throughout the network~:
\[
P_{Local} = \underset{u \in V}{\operatorname{E}} \left[ P_{Local-Adopt}(a,u,t,\Delta_{t})\right]
\]
\end{definition}

\begin{definition}
Let us denote by $P_{Trend} (\Delta_{t}, \frac{V_{a}(t)}{n}, \varepsilon)$ the probability that at time $t + \Delta_{t}$ the group of network members that advocate the trend $a$ has at least $\varepsilon \cdot n$ members (namely, that $|V_{a}(t + \Delta_{t})| \geq \varepsilon \cdot n$).
\end{definition}

We assume that the seed group of members that advocate a trend at time $t$ is randomly placed in the network. Under this assumption we can now present the main result of this work : the lower bound over the prevalence of an emerging trend. Note that we use $P_{Local-Adopt}$ as a modular function in order to allow future validation in other environments. The explicit result is \mbox{given in Theorem \ref{theorem.random.funfmodel}.}

\begin{theorem}
\label{theorem.random.generic}
For any value of $\Delta_{t}$, $|V_{a}(t)|$, $n$, $\varepsilon$, the probability that at time $t + \Delta_{t}$ at least $\epsilon$ portion of the network's users would advocate trend $a$ is lower bounded as follows~:
\remove{
\[
P_{Trend} \left(\Delta_{t}, \frac{|V_{a}(t)|}{n}, \varepsilon\right) \leq  \left(E_{u \in V} \left[ P_{Local-Adopt}(a,u,t,\Delta_{t})\right]\right)^{\varepsilon \cdot n} \cdot
\left(1 - \Phi \left( \sqrt{n} \cdot \frac{\varepsilon - \tilde{P_{+}}}{\sqrt{\tilde{P_{+}} (1-\tilde{P_{+}})    }} \right)\right)
\]
}
\[
P_{Trend} \left(\Delta_{t}, \frac{|V_{a}(t)|}{n}, \varepsilon\right) \geq
P_{Local}^{\varepsilon \cdot n} \cdot
\left(1 - \Phi \left( \frac{\sqrt{n} \cdot (\varepsilon - \tilde{P_{-}})}{\sqrt{\tilde{P_{-}} (1-\tilde{P_{-}})    }} \right)\right)
\]
where~:
\remove{
\[
\tilde{P_{+}} =
  e^{(\rho_{opt_{+}} - \Delta_{t} \cdot \sigma_{+} )} \cdot \left( \frac{\Delta_{t} \cdot \sigma_{+}}{\rho_{opt_{+}}} \right)^{\rho_{opt_{+}}}
\]
}
\[
\tilde{P_{-}} =
    e^{-\left(\frac{\Delta_{t} \cdot \sigma_{-}} {2} -\rho_{opt_{-}}  + \frac{\rho_{opt_{-}}^{2}}{2 \Delta_{t} \cdot \sigma_{-}}\right)}
\]
and where~:
\[
\rho_{opt_{-}} \triangleq \underset{\rho}{\operatorname{argmin}} \ \left(P_{Local}^{\varepsilon \cdot n} \cdot P_{Trend} \left(\Delta_{t}, \frac{|V_{a}(t)|}{n}, \varepsilon\right)\right)
\]
and provided that~:
\remove{
\[
\rho_{opt_{+}} > \Delta_{t} \cdot \sigma_{+}
\]
}
\[
\rho_{opt_{-}} < \Delta_{t} \cdot \sigma_{-}
\]
and as $\sigma_{-}$ depends on $P_{\Delta}$, using the following bound~:
\[
\forall v,u \in V \quad , \quad P_{\Delta} \leq  \frac{c^{2} \cdot  \Delta^{1-\gamma}}{2\gamma^{2}-3\gamma+1}
\]
\begin{proof}

See Appendix for the complete proof of the Theorem

\end{proof}
\end{theorem}

%

Recent studies examined the way influence is being conveyed through social links. In \cite{funfaaai} the probability of network users to install applications, after being exposed to the applications installed by the friends, was tested. This behavior was shown to be best modeled as follows, for some user $v$~:
\begin{equation}
\label{eq.funf1}
P_{Local-Adopt}(a,v,t,\Delta_{t}) = 1 - e^{-(s_{v} + p_{a}(v))}
\end{equation}

Exact definitions and methods of obtaining the values of $s_{v}$ and $w_{v,u}$ can be found in \cite{funfaaai}.
The intuition of these network properties is as follows~:

For every member $v \in V$, $s_{v} \geq 0$ captures the individual susceptibility of this member, regardless of the specific behavior (or trend) in question.
$p_{a}(v)$ denotes the \emph{network potential} for the user $v$ with respect to the trend $a$, and is defined as the sum of network agnostic ``\emph{social weights}'' of the user $v$ with the friends exposing him with the trend $a$.

Notice also that both properties are trend-agnostic. However, while $s_{v}$ is evaluated once for each user and is network agnostic, $p_{a}(v)$ contributes network specific information and can also be used by us to decide the identity of the network's members that we would target in our initial campaign.

Using Theorem \ref{theorem.random.generic} we can now construct a lower bound for the success probability of a campaign, regardless of the specific value of $\rho$~:

\begin{theorem}
\label{theorem.random.funfmodel}
For every $\Delta_{t}$, $|V_{a}(t)|$, $n$, $\varepsilon$, the probability that at time $t + \Delta_{t}$ at least $\epsilon$ portion of the network's users advocate the trend $a$ is~:
\remove{
\[
P_{Trend} \left(\Delta_{t}, \frac{|V_{a}(t)|}{n}, \varepsilon\right) \leq  e^{-\varepsilon \cdot n \cdot \xi_{G} \cdot \xi_{N}^{\rho_{opt_{+}}}} \cdot
\left(1 - \Phi \left( \sqrt{n} \cdot \frac{\varepsilon - \tilde{P_{+}}}{\sqrt{\tilde{P_{+}} (1-\tilde{P_{+}})    }} \right)\right)
\]
}
\[
P_{Trend} \left(\Delta_{t}, \frac{|V_{a}(t)|}{n}, \varepsilon\right) \geq
e^{-\varepsilon \cdot n \cdot \xi_{G} \cdot \xi_{N}^{\rho_{opt_{-}}}} \cdot
\left(1 - \Phi \left( \sqrt{n} \cdot \frac{\varepsilon - \tilde{P_{-}}}{\sqrt{\tilde{P_{-}} (1-\tilde{P_{-}})    }} \right)\right)
\]

where~:
\remove{
\[
\tilde{P_{+}} =
  e^{(\rho_{opt_{+}} - \Delta_{t} \cdot \sigma_{+} )} \cdot \left( \frac{\Delta_{t} \cdot \sigma_{+}}{\rho_{opt_{+}}} \right)^{\rho_{opt_{+}}}
\]
}
\[
\tilde{P_{-}} =
    e^{-(\frac{\Delta_{t} \cdot \sigma_{-}} {2} -\rho_{opt_{-}}  + \frac{\rho_{opt_{-}}^{2}}{2 \Delta_{t} \cdot \sigma_{-}})}
\]
and where~:
\remove{
\[
\rho_{opt_{+}} \triangleq \underset{\rho}{\operatorname{argmax}} \ \left(e^{-\varepsilon \cdot n \cdot \xi_{G} \cdot \xi_{N}^{\rho}} \cdot P_{Trend} \left(\Delta_{t}, \frac{|V_{a}(t)|}{n}, \varepsilon\right)\right)
\]
}
\[
\rho_{opt_{-}} \triangleq \underset{\rho}{\operatorname{argmin}} \ \left(e^{-\varepsilon \cdot n \cdot \xi_{G} \cdot \xi_{N}^{\rho}} \cdot P_{Trend} \left(\Delta_{t}, \frac{|V_{a}(t)|}{n}, \varepsilon\right)\right)
\]
and provided that~:
\remove{
\[
\rho_{opt_{+}} > \Delta_{t} \cdot \sigma_{+}
\]
}
\[
\rho_{opt_{-}} < \Delta_{t} \cdot \sigma_{-}
\]

and where $\xi_{G}$ denotes the network's \emph{adoption factor} and $\xi_{N}$ denotes the network's \emph{influence factor}~:
\[
\xi_{G} = e^{-\frac{1}{n}\sum_{v \in V}s_{v}}
\qquad , \qquad
\xi_{N} = e^{ - \frac{1}{n} \sum_{e(v,u) \in E} (\frac{w_{u,v}}{|\mathcal{N}_{v}|} + \frac{w_{v,u}}{|\mathcal{N}_{u}|})}
\]
\begin{proof}
See complete proof in the Appendix.
\end{proof}
\end{theorem}

\section{Experimental Results}
\label{sec.results}

We have validated our model using two comprehensive datasets, the \emph{Friends and Family} dataset that studied the casual and social aspects of a small community of students and their friends in Cambridge, and the \emph{eToro} dataset --- the entire financial transactions of over 1.5M users of a ``social trading;; community.

The datasets were analyzed using the model given in \cite{funfaaai}, based on which we have experimentally calculated the values of $\beta, \xi_{G}, \xi_{N}$ and $\sigma_{-}$.

\begin{figure}[htb]
   \centering
   \includegraphics[width=0.47 \textwidth,keepaspectratio]{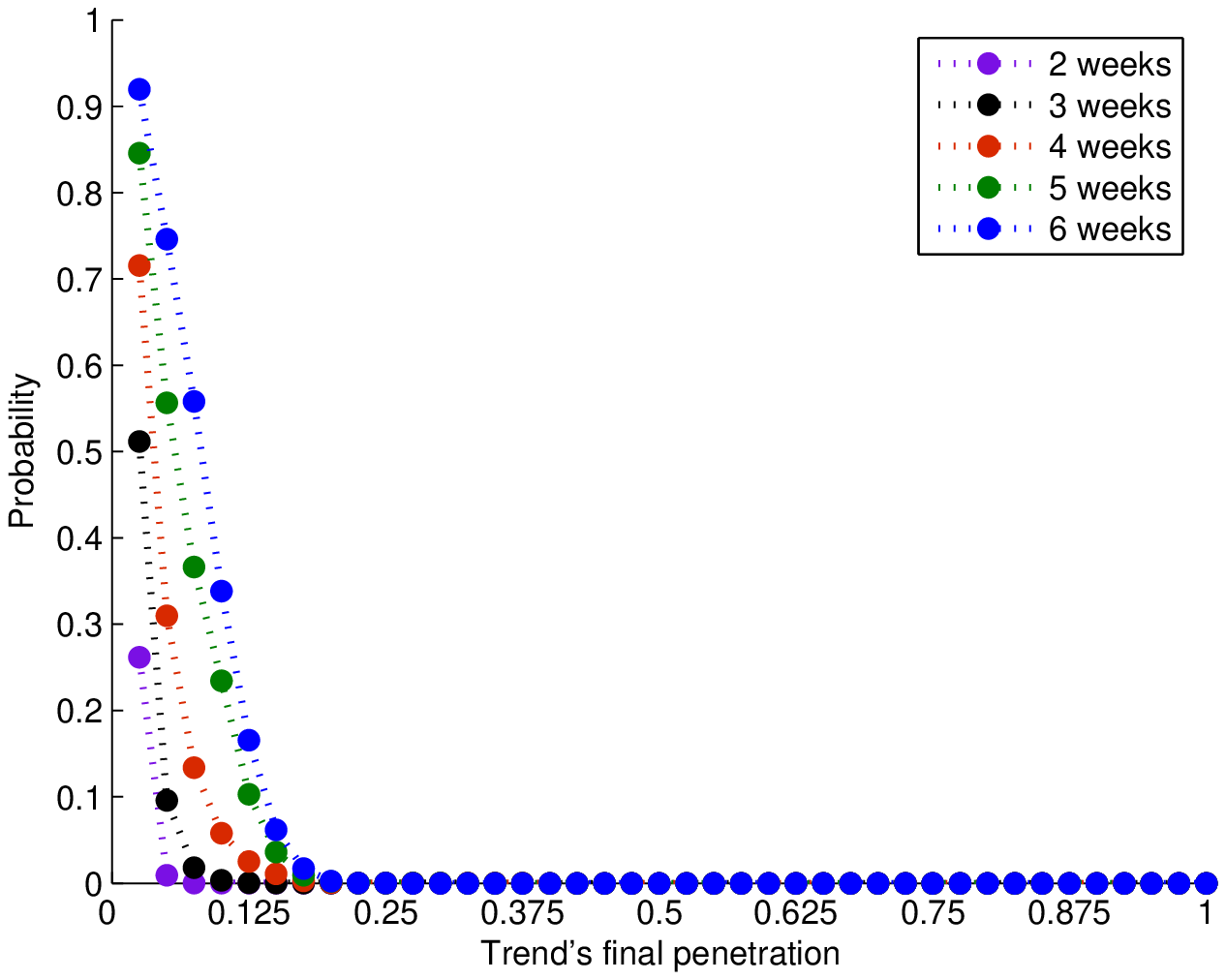}
   \includegraphics[width=0.47 \textwidth,keepaspectratio]{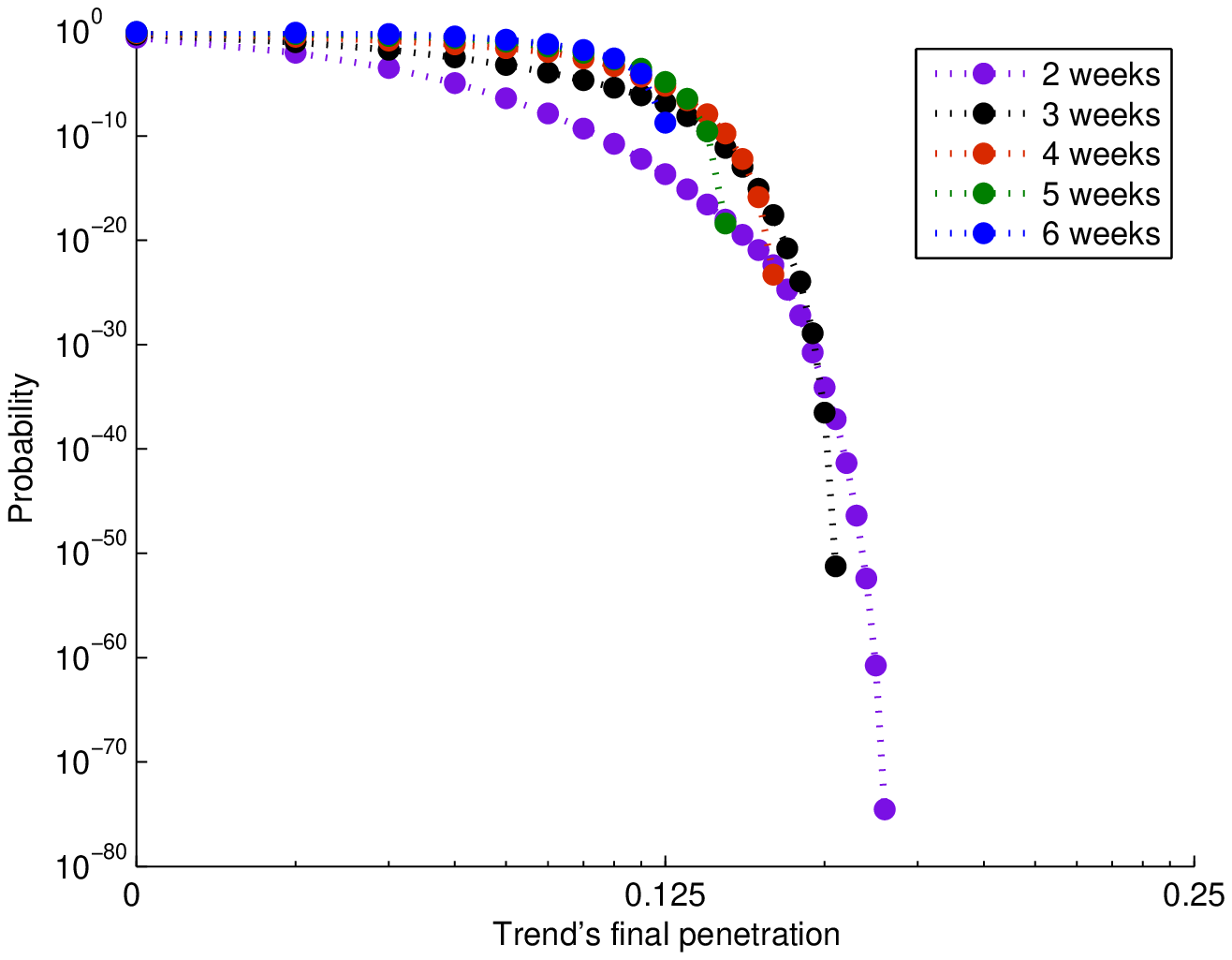}
   \caption{Trends spreading potential in the \emph{eToro} network, for various penetration rates. Initial seed group is defined as 5\% of the population. Each curve represents a different time period, from 2 weeks to 6 weeks.
   }
   \label{fig.etoro1}
\end{figure}

\begin{figure}[htb]
   \centering
   \includegraphics[width=0.48 \textwidth,keepaspectratio]{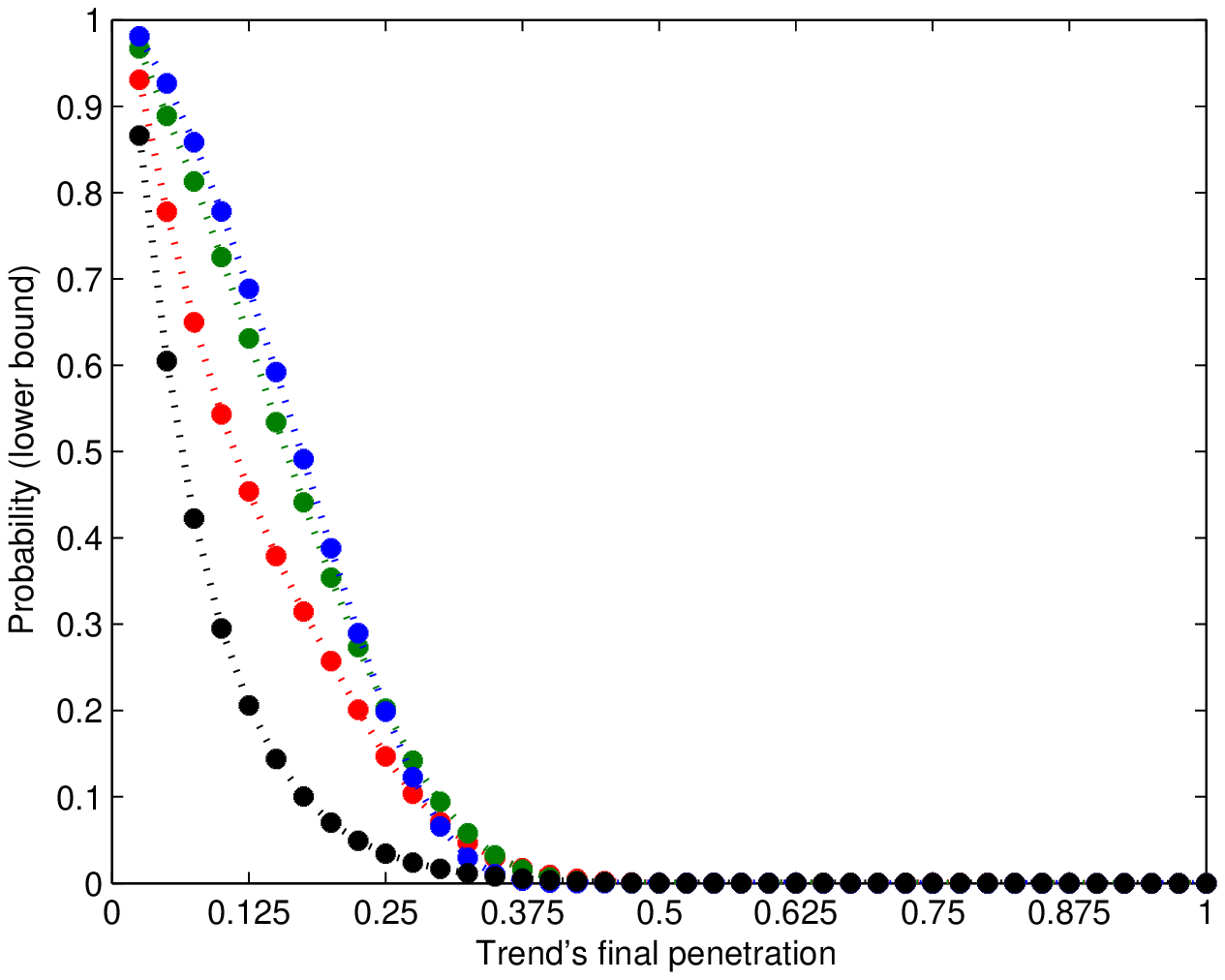}
   \includegraphics[width=0.48 \textwidth,keepaspectratio]{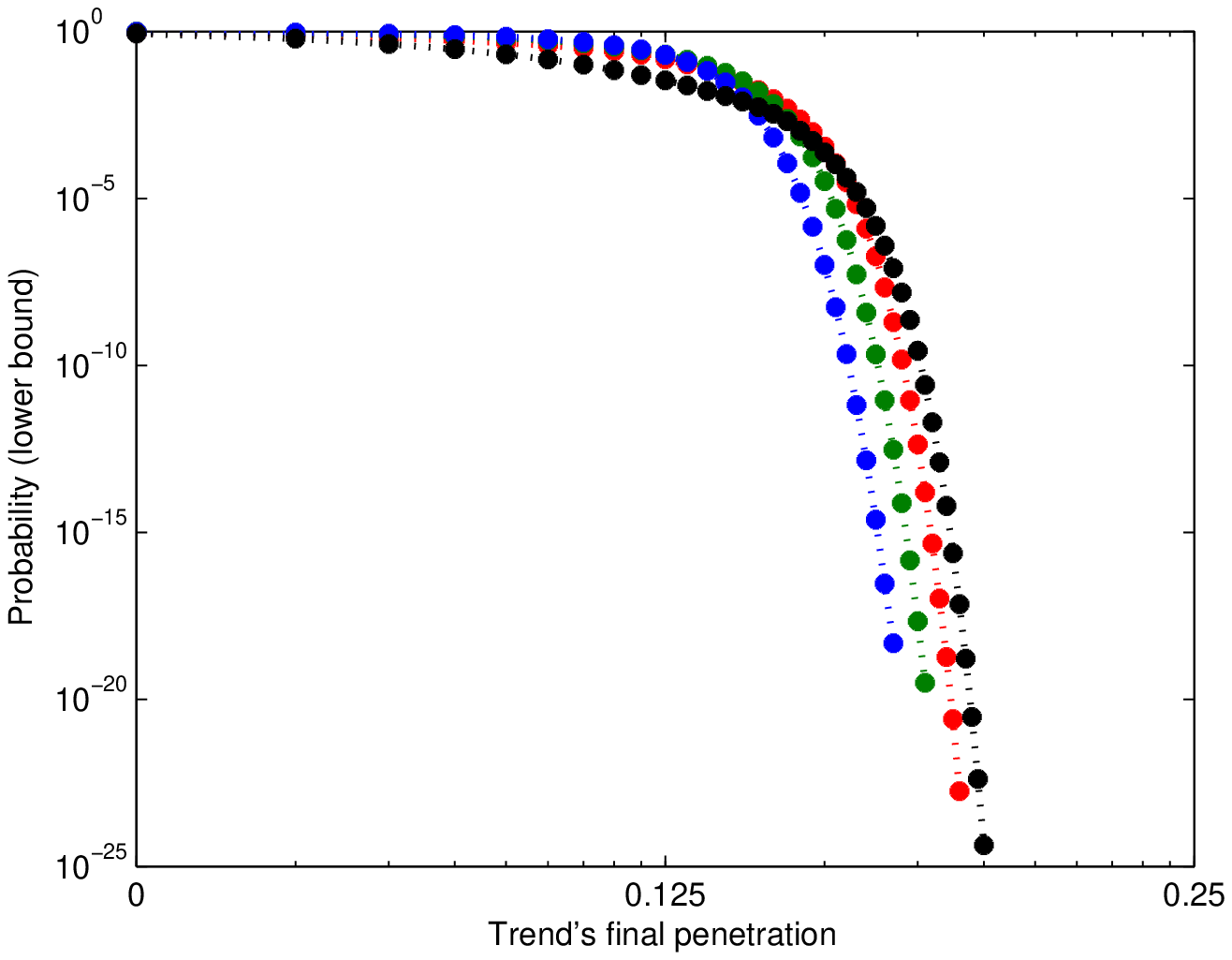}
   \caption{Trends spreading potential in the \emph{Friends and Family} network, for various penetration rates. Initial seed group is defined as 5\% of the population. Each curve represents a different time period, from 2 weeks to 5 weeks.
   }
   \label{fig.funf1}
\end{figure}

Figures \ref{fig.etoro1} and \ref{fig.funf1}  demonstrate the probabilistic lower bound for trend emergence, as a function of the overall penetration of the trend at the end of the time period, under the assumption that the emerging trend was observed in 5\% of the population. In other words, for any given ``magnitude'' of trends, what is the probability that network phenomena that are being advocated by 5\% of the network, would spread to this magnitude.
Notice that although a longer spreading time slightly improve the penetration probability, the ``maximal outreach'' of the trend (the maximal rate of global adoption, with sufficient probability) is dominated by the topology of the network, and the local adoption features.

\section{Conclusions and Future Work}
\label{sec.conc}

In this work we have discussed the problem of trends prediction, that is --- observing anomelous network patterns and predicting which of them would become a prominent trend, spreading successfully throughout the network. We have analyzed this problem using information diffusion techniques, and have presented a lower bound for the probability of a pattern to become a global trend in the network, for any desired level of spreading.
In order to model the local interaction between members, we have used the results from \cite{funfaaai} that studied the local social influence dynamics between members of social networks.
\remove{
It is also interesting to mentioned that the influence model of \cite{funfaaai} is also a good approximated of the \emph{Gompertz Function} --- a model that is frequently used for predicting the dynamics of a great variety of processes, such as mobile phone uptake \cite{Gompertz-mobile}, population in a confined space \cite{Gompertz-population}, or growth of tumors \cite{Gompertz-tumors}.
}

Though our work provides a comprehensive theoretical framework to understand trends diffusion in social networks, there are still a few challenges ahead. For example, we wish to extend our model to other network models such as Erdos-Renyi random networks, as well as Small World networks. This is essential as more evidences are suggesting that some communities involve complex structures that cannot be easily approximated using s simplistic scale-free model~\cite{leskovec2005graphs}.

In addition, our results can be used in order to provide answers to other questions, such as what is the optimal group of members that should be used as a ``seed group'' in order to maximize the effects of marketing campaigns. Another example might be finding changes in the topology of the social network that would influence the information diffusion progress in a desired way (either to encourage or surpass certain emerging trends).

In order to achieve these goals we are planning a large-scale field test with a leading online social platform, that would give us access to collect more empirical supporting evidences, as well as conducting an active experiment in which we would try to predict trends in real time.

Finally, we are interested in comparing the prediction obtained from our model with the actual semantics of the trends, to better understand the connection between the trends semantics and the diffusion process they undergo.

\bibliographystyle{elsarticle-num}

\bibliography{/Altshuler}

\newpage
\appendix{\huge{\textbf{Appendix}}}

\section{Proof of Theorem 1}
\label{appendix.a}

In order to prove Theorem 1 we shall require the following definitions~:

\begin{definition}
Let $\mathcal{N}_{v,a}(t)$ denote the number of friends of user $v$ that at time $t$ are exposing $v$ to the trend $a$ (namely, the number of friends of $v$ that at time $t$ have been exposed to the trend $a$ and are conveying this information to $v$).
\end{definition}

Note that ``exposing'' a neighbor to a trend does not necessarily mean advocating the same trend.

\begin{definition}
Let us denote by $P_{\rho-Trend} (\Delta_{t}, \rho, \frac{|V_{a}(t)|}{n}, \varepsilon)$ the probability that at time $t + \Delta_{t}$ at least $\varepsilon \cdot n$ members of the network have been exposed to the trend $a$ by at least $\rho$ of their friends.
\end{definition}

In addition, we define $\rho_{opt_{-}}$ that is used in the Theorem~:

\begin{definition}
\remove{
\[
\rho_{opt_{+}} \triangleq \underset{\rho}{\operatorname{argmax}} \ \left(\left(\underset{u \in V}{\operatorname{E}} \left[ P_{Local-Adopt}(a,u,t,\Delta_{t})\right]\right)^{\varepsilon \cdot n} \cdot P_{Trend} \left(\Delta_{t}, \frac{|V_{a}(t)|}{n}, \varepsilon\right)\right)
\]
}
\[
\rho_{opt_{-}} \triangleq \underset{\rho}{\operatorname{argmin}} \ \left(P_{Local}^{\varepsilon \cdot n} \cdot P_{Trend} \left(\Delta_{t}, \frac{|V_{a}(t)|}{n}, \varepsilon\right)\right)
\]

We later see that the expression for $P_{Trend}$ would refer to $\rho$. Using $\rho_{Opt_{-}}$ we would later be able to omit this dependance.
\end{definition}

\noindent\textbf{Theorem 1. }

For any value of $\Delta_{t}$, $|V_{a}(t)|$, $n$, $\varepsilon$, the probability that at time $t + \Delta_{t}$ at least $\epsilon$ portion of the network's users would advocate trend $a$ is lower bounded as follows~:
\remove{
\[
P_{Trend} \left(\Delta_{t}, \frac{|V_{a}(t)|}{n}, \varepsilon\right) \leq  \left(E_{u \in V} \left[ P_{Local-Adopt}(a,u,t,\Delta_{t})\right]\right)^{\varepsilon \cdot n} \cdot
\left(1 - \Phi \left( \sqrt{n} \cdot \frac{\varepsilon - \tilde{P_{+}}}{\sqrt{\tilde{P_{+}} (1-\tilde{P_{+}})    }} \right)\right)
\]
}
\[
P_{Trend} \left(\Delta_{t}, \frac{|V_{a}(t)|}{n}, \varepsilon\right) \geq
P_{Local}^{\varepsilon \cdot n} \cdot
\left(1 - \Phi \left( \frac{\sqrt{n} \cdot (\varepsilon - \tilde{P_{-}})}{\sqrt{\tilde{P_{-}} (1-\tilde{P_{-}})    }} \right)\right)
\]
where~:
\remove{
\[
\tilde{P_{+}} =
  e^{(\rho_{opt_{+}} - \Delta_{t} \cdot \sigma_{+} )} \cdot \left( \frac{\Delta_{t} \cdot \sigma_{+}}{\rho_{opt_{+}}} \right)^{\rho_{opt_{+}}}
\]
}
\[
\tilde{P_{-}} =
    e^{-\left(\frac{\Delta_{t} \cdot \sigma_{-}} {2} -\rho_{opt_{-}}  + \frac{\rho_{opt_{-}}^{2}}{2 \Delta_{t} \cdot \sigma_{-}}\right)}
\]
and provided that~:
\remove{
\[
\rho_{opt_{+}} > \Delta_{t} \cdot \sigma_{+}
\]
}
\[
\rho_{opt_{-}} < \Delta_{t} \cdot \sigma_{-}
\]
\begin{proof}

We first assess the number of ``agents'' residing in adjacent vertices from some vertex $v$ at any given time~:
\begin{lemma}\label{lemma.number.scalefree}
Let $v \in V$ be an arbitrary user of the network $G$. Then~:
\[
P\left[E_{v}[\mathcal{N}_{v,a}(t + \Delta_{t}) - \mathcal{N}_{v,a}(t)] = \eta \cdot \beta^{\Delta_{t}} \cdot \frac{|V_{a}(t)|}{n} \right] = c \cdot \eta^{-\gamma}
\]
\begin{proof}
We assume that the movement of the agents in the network are random\footnote{People may sustain different kinds of influence from different members of their immediate social network. However, once users had already adopted a trend, the source of influence that caused their behavioral change has no effect on way they would later affect their friends}. Hence~:
\[\forall u \in V(G) \ , \ E[\emph{number of agents residing on } u] = \frac{\# \emph{of agents}}{n}
\]

At time $t$ there are $|V_{a}(t)|$ members of the network that advocate trend $a$. Those members generate on average $\beta$ ``agents'' that are sent along the social links to their friends, creating chains of length $\Delta_{t}$, and a total of $\beta^{\Delta_{t}} \cdot |V_{a}(t)|$ active agents.
Incorporating this with the distribution of the degrees, the rest is implied.
\end{proof}
\end{lemma}

The following Lemma produces the probability that two arbitrarily selected vertices would have degrees which differ in more than a certain threshold~:
\begin{lemma}
\label{lemma.degreeratio}
\[
\forall v,u \in V \quad , \quad P_{\Delta} \leq  \frac{c^{2} \cdot  \Delta^{1-\gamma}}{2\gamma^{2}-3\gamma+1}
\]
\begin{proof}
By definition~:
\[
P_{\Delta} = Prob \left[ deg(u) \geq \Delta \cdot deg(v) \right] \leq
\]
\[
\int_{1}^{\infty} \int_{\Delta}^{\infty}\left(Prob\left[deg(v) = j\right] \cdot Prob\left[deg(u) = m \cdot j\right] dm\right)dj
\]

As the network is scale free, we can write~:
\[
P_{\Delta} = Prob \left[ deg(u) \geq \Delta \cdot deg(v) \right] \leq
\]
\[
\int_{1}^{\infty} \int_{\Delta}^{\infty}\left(c \cdot j^{-\gamma} \cdot c \cdot (m \cdot j)^{-\gamma} dm\right)dj \leq
\]
\[
\int_{1}^{\infty} \left( \left. \frac{c^{2} \cdot j^{-2\gamma} \cdot m^{-\gamma+1}}{1-\gamma} \right|_{m=\Delta}^{\infty} \right)dj \leq
\]
\[
- \int_{1}^{\infty} \frac{c^{2} \cdot j^{-2\gamma} \cdot \Delta^{-\gamma+1}}{1-\gamma} dj \leq
- \left. \frac{c^{2} \cdot j^{-2\gamma+1} \cdot \Delta^{1-\gamma}}{(1-2\gamma)(1-\gamma)} \right|_{j=1}^{\infty} \leq
\]
\[
\frac{c^{2} \cdot  \Delta^{1-\gamma}}{(2\gamma-1)(\gamma-1)} \leq
\frac{c^{2} \cdot  \Delta^{1-\gamma}}{2\gamma^{2}-3\gamma+1}
\]
\end{proof}
\end{lemma}

\begin{lemma}
\label{lemma.prob2}
For any member $v \in V$ at time $t + \Delta_{t}$, the probability that $v$ will be exposed at
the next time-step to the trend $a$ is at least $\sigma_{-}$. \remove{and at most $\sigma_{+}$.}
\begin{proof}
The probability that an agent located on a vertex $u$ such that $(u,v) \in E$ will move to $v$ at the next time-step is $\frac{1}{deg(u)}$. Therefore, remembering that $v$ has $\mathcal{N}_{v,a}(t)$ agents that resides in neighboring vertices in time $t$, the probability that $v$ will \emph{not} be exposed to the trend at the next time-step is~:
\begin{equation}
\label{eq.prob1}
\left(1-\frac{1}{deg(u)}\right)^{\mathcal{N}_{v,a}(t)}
\end{equation}

Using the well known inequality $(1-x) < e^{-x}$ for $x < 1$, and taking into account Lemma \ref{lemma.number.scalefree}, we can bound Equation \ref{eq.prob1} from above by~:
\begin{equation}
\label{eq.prob2}
\left(1-\frac{1}{deg(u)}\right)^{\mathcal{N}_{v,a}(t)}  \leq
e^{-\frac{\mathcal{N}_{v,a}(t)}{deg(u)}} \leq
e^{-\frac{deg(v)}{deg(u)} \cdot \frac{\beta^{\Delta_{t}} \cdot |V_{a}(t)|}{n}}
\end{equation}

Using Lemma \ref{lemma.degreeratio} we can further simplify Equation \ref{eq.prob2} as follows~:
\begin{equation}
\label{eq.prob3}
Prob\left[e^{-\frac{deg(v)}{deg(u)} \cdot \frac{\beta^{\Delta_{t}} \cdot |V_{a}(t)|}{n}} \leq e^{-\Delta \cdot \frac{\beta^{\Delta_{t}} \cdot |V_{a}(t)|}{n}}\right] \geq
(1 - P_{\Delta})
\end{equation}
Therefore, the probability that a user will be exposed to the trend in the next time step is at least~:
\[
1 - e^{-\Delta \cdot \frac{\beta^{\Delta_{t}} \cdot |V_{a}(t)|}{n}} \cdot (1 - P_{\Delta})
\]
which equals $\sigma_{-}$.

\end{proof}
\end{lemma}

We can now proceed to the calculation of $P_{\rho-Trend}$.

\begin{lemma}
\label{theorem.robust3}
The probability that at time $t + \Delta_{t}$ at least $\varepsilon \cdot n$ members of the network have been exposed to a trend $a$ by at least $\rho$ of their friends is lower bounded as follows~:
\[
P_{\rho-Trend} (\Delta_{t}, \rho, \frac{|V_{a}(t)|}{n}, \varepsilon) \geq  1 - \Phi \left( \sqrt{n} \cdot \frac{\varepsilon - \tilde{P_{-}}}{\sqrt{\tilde{P_{-}} (1-\tilde{P_{-}})    }} \right)
\]
where~:
\[
\tilde{P_{-}} =
    e^{-(\frac{\Delta_{t} \cdot \sigma_{-}} {2} -\rho  + \frac{\rho^{2}}{2 \Delta_{t} \cdot \sigma_{-}})}
\]
and where $\Phi(x)$ is the cumulative normal distribution function, defined as~:
\[
\Phi(x) = \frac{1}{\sqrt{2 \pi}} \int_{-\infty}^{x} e^{-\frac{1}{2}t^{2}} dt
\]
and also provided that~:
\[
\rho < \Delta_{t} \cdot \sigma_{-}
\]
\begin{proof}
Using Lemma~\ref{lemma.prob2} we have a lower bound for the probability that a user $v$ will be exposed to some trend $a$ by an agent originated by one of the group of users that advocate the trend $a$ at time $t$. This is in fact a \emph{Bernoulli} trial with success probability of
$\sigma_{-}$.

Denoting $X_v(t)$ the number of times user  $v$ is being exposed to the trend $a$ after $t$ steps, we shall now use the negative variance \emph{Chernoff} bound~:

\[ P[X_v(t) < (1-\delta) t \cdot \sigma_{-}] < e^{-\delta^2 \frac{t \cdot \sigma_{-}} {2}}\]

Once selecting $\delta = 1 - \frac{\rho}{t \cdot \sigma_{-}}$ and for the entire lifespan of the trend (namely, for $t = \Delta_{T}$) we obtain the probability that a single (specific) member will be exposed to the trend $a$ at least $\rho$ times. For this, we shall first define~:
\[
  \tilde{P_{-}} \triangleq P_{\rho-Trend} (\Delta_{T}, \rho, \frac{|V_{a}(t)|}{n}, n^{-1})
\]
which by definition implies~:
\[
\tilde{P_{-}}  =   P[X_v(\Delta_{T}) < \rho]
 < e^{-(1-\frac{\rho}{\Delta_{T} \cdot \sigma_{-}})^2 \frac{\Delta_{T} \cdot \sigma_{-}} {2}} <
\]
\[
e^{-(1 -2\frac{\rho}{\Delta_{T} \cdot \sigma_{-}} + \frac{\rho}{\Delta_{T} \cdot \sigma_{-}}^{2}) \frac{\Delta_{T} \cdot \sigma_{-}} {2}} <
e^{-(\frac{\Delta_{T} \cdot \sigma_{-}} {2} -\rho  + \frac{\rho^{2}}{2 \Delta_{T} \cdot \sigma_{-}})}
\]

As the \emph{Chernoff} bound requires that $\delta > 0$ we should make sure that~:
\[
\rho < \Delta_{T} \cdot \sigma_{-}
\]

As we want to bound the probability that at least $\varepsilon \cdot n$ of the network members are exposed to the trend at least $\rho$ times, we shall use the above as a success probability of yet a second \emph{Bernoulli} trial. As $n$ is assumed to be large, the number of exposed members can be approximated using \emph{Normal} distribution~:
\[
P_{\rho-Trend} \left(\Delta_{T}, \rho, \frac{|V_{a}(t)|}{n}, \varepsilon\right) \geq  1 - \Phi \left( \frac{\varepsilon \cdot n - n \cdot \tilde{P_{-}}}{\sqrt{n \cdot \tilde{P_{-}} (1-\tilde{P_{-}})    }} \right)
\]
and the rest is implied.
\end{proof}
\end{lemma}

Whereas Lemma \ref{theorem.robust3} provides an estimation concerning the global outreach of trends in terms of exposure, it does not take into account the probability that users that are exposed to the trend by $\rho$ of their friends, will actually adopt it. In order to do this, we need to incorporate $P_{Local-Adopt}$ into Lemma \remove{\ref{theorem.robust1} and } \ref{theorem.robust3}, producing a combined bound for the global adoption of the trend.

\begin{proposition}
\label{theorem.robust2}
For any $\Delta_{t}$, $|V_{a}(t)|$, $n$, $\varepsilon$, the probability that at time $t + \Delta_{t}$ at least $\epsilon$ portion of the network's users advocate the legitimate trend $a$ is~:
\remove{
\[
P_{Trend,\rho} \left(\Delta_{t}, \rho, \frac{|V_{a}(t)|}{n}, \varepsilon\right) =  P_{Local-Adopt}^{\varepsilon \cdot n} \cdot P_{\rho-Trend} \left(\Delta_{t}, \rho, \frac{|V_{a}(t)|}{n}, \varepsilon\right)
\]
}
\[
P_{Trend,\rho} \left(\Delta_{t}, \rho, \frac{|V_{a}(t)|}{n}, \varepsilon\right) =  P_{Local}^{\varepsilon \cdot n} \cdot P_{\rho-Trend} \left(\Delta_{t}, \rho, \frac{|V_{a}(t)|}{n}, \varepsilon\right)
\]
\end{proposition}

Notice that $\rho$ appears in the expression of Proposition \ref{theorem.robust2} for mathematical reasons, and has no actual meaning. We omit the dependency of the expression on $\rho$, by finding the optimal value of $\rho$ for every set of values of $\varepsilon$, $|V_{a}(t)|$ and $\Delta_{t}$, by assigning~:
\[
P_{Trend} \left(\Delta_{t}, \frac{|V_{a}(t)|}{n}, \varepsilon\right) =
P_{Trend,\rho} \left(\Delta_{t}, \rho_{opt_{-}}, \frac{|V_{a}(t)|}{n}, \varepsilon\right)
\]

And the rest is implied.
\end{proof}

\section{Proof of Theorem 2}
\label{appendix.b}

Let us remind once again the local influence model that was shown in \cite{funfaaai} to best approximate the behavior diffusion in real world social networks~:
\begin{equation}
\label{eq.funf1}
P_{Local-Adopt}(a,v,t,\Delta_{t}) = 1 - e^{-(s_{v} + p_{a}(v))}
\end{equation}

We recall that $s_{v} \geq 0$ captures the individual susceptibility of this member, and that
$p_{a}(v)$ denotes the \emph{network potential} for the user $v$ with respect to the trend $a$, and is defined as the sum of network agnostic ``\emph{social weights}'' of the user $v$ with the friends exposing him with the trend $a$~:
\[
p_{a}(v) = \sum_{u \in \mathcal{N}_{v,a}} w_{v,u}
\]
(where $\mathcal{N}_{v,a}$ is the overall group of users exposing $v$ to the trend $a$).

Using Theorem \ref{theorem.random.generic} we can now construct a lower bound for the success probability of a campaign, regardless of the specific value of $\rho$~:

\noindent\textbf{Theorem 2.}
For every $\Delta_{t}$, $|V_{a}(t)|$, $n$, $\varepsilon$, the probability that at time $t + \Delta_{t}$ at least $\epsilon$ portion of the network's users advocate the legitimate trend $a$ is~:
\remove{
\[
P_{Trend} \left(\Delta_{t}, \frac{|V_{a}(t)|}{n}, \varepsilon\right) \leq  e^{-\varepsilon \cdot n \cdot \xi_{G} \cdot \xi_{N}^{\rho_{opt_{+}}}} \cdot
\left(1 - \Phi \left( \sqrt{n} \cdot \frac{\varepsilon - \tilde{P_{+}}}{\sqrt{\tilde{P_{+}} (1-\tilde{P_{+}})    }} \right)\right)
\]
}
\[
P_{Trend} \left(\Delta_{t}, \frac{|V_{a}(t)|}{n}, \varepsilon\right) \geq
e^{-\varepsilon \cdot n \cdot \xi_{G} \cdot \xi_{N}^{\rho_{opt_{-}}}} \cdot
\left(1 - \Phi \left( \sqrt{n} \cdot \frac{\varepsilon - \tilde{P_{-}}}{\sqrt{\tilde{P_{-}} (1-\tilde{P_{-}})    }} \right)\right)
\]

where~:
\remove{
\[
\tilde{P_{+}} =
  e^{(\rho_{opt_{+}} - \Delta_{t} \cdot \sigma_{+} )} \cdot \left( \frac{\Delta_{t} \cdot \sigma_{+}}{\rho_{opt_{+}}} \right)^{\rho_{opt_{+}}}
\]
}
\[
\tilde{P_{-}} =
    e^{-(\frac{\Delta_{t} \cdot \sigma_{-}} {2} -\rho_{opt_{-}}  + \frac{\rho_{opt_{-}}^{2}}{2 \Delta_{t} \cdot \sigma_{-}})}
\]
and where~:
\remove{
\[
\rho_{opt_{+}} \triangleq \underset{\rho}{\operatorname{argmax}} \ \left(e^{-\varepsilon \cdot n \cdot \xi_{G} \cdot \xi_{N}^{\rho}} \cdot P_{Trend} \left(\Delta_{t}, \frac{|V_{a}(t)|}{n}, \varepsilon\right)\right)
\]
}
\[
\rho_{opt_{-}} \triangleq \underset{\rho}{\operatorname{argmin}} \ \left(e^{-\varepsilon \cdot n \cdot \xi_{G} \cdot \xi_{N}^{\rho}} \cdot P_{Trend} \left(\Delta_{t}, \frac{|V_{a}(t)|}{n}, \varepsilon\right)\right)
\]
and provided that~:
\remove{
\[
\rho_{opt_{+}} > \Delta_{t} \cdot \sigma_{+}
\]
}
\[
\rho_{opt_{-}} < \Delta_{t} \cdot \sigma_{-}
\]

and where $\xi_{G}$ denotes the network's \emph{adoption factor} and $\xi_{N}$ denotes the network's \emph{influence factor}~:
\[
\xi_{G} = e^{-\frac{1}{n}\sum_{v \in V}s_{v}}
\]
\[
\xi_{N} = e^{ - \frac{1}{n} \sum_{e(v,u) \in E} (\frac{w_{u,v}}{|\mathcal{N}_{v,a}|} + \frac{w_{v,u}}{|\mathcal{N}_{u,a}|})}
\]
\begin{proof}
From Equation \ref{eq.funf1} we have~:
\[
P_{Local-Adopt}(a,v,t,\Delta_{t}) = 1 - e^{-(s_{v} + p_{a}(v))}
\]

\remove{
Notice that the values of $P_{Local-Adopt}(a,v)$ for different users may be very different. Subsequently, the product of $\varepsilon \cdot n$ local adoption probabilities may be very different for different groups of $\varepsilon \cdot n$ users. As we are interested in an upper bound, we shall assume that the distribution of $P_{Local-Adopt}(v)$ among users is one that maximize the expected product of $P_{Local-Adopt}(v)$ for a random group of $\varepsilon \cdot n$ users. This distribution is the uniform distribution that equals the expected value of $P_{Local-Adopt}$~:
}

The expected value of the local adoption probability is therefore~:
\[
E_{u \in V} [P_{Local-Adopt}(u)] = \frac{1}{n} \sum_{v \in V} 1 - e^{-(s_{v} + \frac{\rho}{|\mathcal{N}_{v,a}|}\sum_{u \in \mathcal{N}_{v,a}} w_{v,u})}
\]
(where $\mathcal{N}_{v,a}$ is the groups of user $v$'s friends).

Using the inequality $(1-x) < e^{-x}$ for $x < 1$, we see that~:
\begin{equation}
\label{eq.local.adopt1}
P_{Local-Adopt}^{\varepsilon \cdot n} = \left(\frac{1}{n} \sum_{v \in V} 1 - e^{-(s_{v} + \frac{\rho}{|\mathcal{N}_{v,a}|}\sum_{u \in \mathcal{N}_{v,a}} w_{v,u})}\right)^{\varepsilon \cdot n} =
\end{equation}
\[
\left(1 - \frac{1}{n} \sum_{v \in V} e^{-(s_{v} + \frac{\rho}{|\mathcal{N}_{v,a}|}\sum_{u \in \mathcal{N}_{v,a}} w_{v,u})}\right)^{\varepsilon \cdot n} < \]
\[
e^{-\varepsilon \cdot \sum_{v \in V} e^{-(s_{v} + \frac{\rho}{|\mathcal{N}_{v,a}|}\sum_{u \in \mathcal{N}_{v,a}} w_{v,u})}}
\]

Using the fact that an arithmetic mean is always greater than a geometric mean, Equation \ref{eq.local.adopt1} can be written as follows~:
\begin{equation}
\label{eq.local.adopt2}
P_{Local-Adopt}^{\varepsilon \cdot n} < e^{-\varepsilon \cdot \sum_{v \in V} e^{-(s_{v} + \frac{\rho}{|\mathcal{N}_{v,a}|}\sum_{u \in \mathcal{N}_{v,a}} w_{v,u})}} <
\end{equation}
\[
e^{-\varepsilon \cdot n \cdot \sqrt[n]{\prod_{v \in V} e^{-(s_{v} + \frac{\rho}{|\mathcal{N}_{v,a}|}\sum_{u \in \mathcal{N}_{v,a}} w_{v,u})}}} <
\]
\[
e^{-\varepsilon \cdot n \cdot e^{-\frac{1}{n}\sum_{v \in V}(s_{v} + \frac{\rho}{|\mathcal{N}_{v,a}|}\sum_{u \in \mathcal{N}_{v,a}} w_{v,u})}} <
\]
\[
e^{-\varepsilon \cdot n \cdot e^{-\frac{1}{n}\sum_{v \in V}s_{v}} \cdot e^{ - \frac{\rho}{n} \sum_{e(v,u) \in E} (\frac{w_{u,v}}{|\mathcal{N}_{v,a}|} + \frac{w_{v,u}}{|\mathcal{N}_{u,a}|})}}
\]

Integrating Equation \ref{eq.local.adopt2} with Proposition \ref{theorem.robust2} and Lemma \ref{theorem.robust3} the rest is implied.
\end{proof}

\end{document}